\documentclass[twoside]{article}
\usepackage{epsfig}

\newcommand{\be}{\begin{equation}}
\newcommand{\ee}{\end{equation}}
\newcommand{\bea}{\begin{eqnarray}}
\newcommand{\eea}{\end{eqnarray}}
\newlength{\hoehe}
\newlength{\tiefe}

\newcommand{\rscp}[2]{\mbox{$\left( #1 ,\, #2 \right)$}}
\newcommand{\Brscp}[2]{\mbox{$\Big( #1 ,\, #2 \Big)$}}

\begin{document}

\title{{\bf Mixtures of Gaussian process priors}\footnote{This is an extended
version of a contribution to the 
Ninth International Conference on Artificial Neural Networks (ICANN 99),
7--10 September 1999, Edinburgh, UK.}}
\author{\large J\"org C.\ Lemm\\
  \normalsize Institut f\"ur Theoretische Physik I, 
              Universit\"at M\"unster\\
  \normalsize D--48149 M\"unster, Germany\\
  {\small E-mail: lemm@uni-muenster.de}\\
  {\small http://pauli.uni-muenster.de/${}^\sim$lemm}\\
  {\small Publication No.: MS-TP1-99-5}
\date{}}
 \maketitle

 \begin{abstract}
\noindent 

{\small
Nonparametric Bayesian approaches
based on Gaussian processes have recently become popular 
in the empirical learning community.
They encompass many classical methods of statistics,
like Radial Basis Functions or various splines,
and are technically convenient because Gaussian integrals
can be calculated analytically.
Restricting to Gaussian processes, however,
forbids for example the implemention of
genuine nonconcave priors.
Mixtures of Gaussian process priors, on the other hand, 
allow the flexible implementation
of complex and situation specific, also nonconcave 
{\it a priori} information.
This is essential for tasks with,
compared to their complexity, 
a small number of available training data.
The paper concentrates on 
the formalism for Gaussian regression problems
where prior mixture models provide a generalisation of 
classical quadratic, typically smoothness related, regularisation
approaches
being more flexible
without having a much larger computational complexity.
}
\end{abstract}
 
\tableofcontents
 
\section{Introduction}

The generalisation behaviour of statistical learning algorithms 
relies essentially on the correctness of the implemented 
{\it a priori} information.
While Gaussian processes and the related regularisation approaches
have, on one hand,
the very important advantage of being able to formulate 
{\it a priori} information
explicitly in terms of the function of interest
(mainly in the form of
smoothness priors which have a long tradition
in density estimation and regression problems
\cite{Whittaker-1923,Wahba-1990,Girosi-Jones-Poggio-1995})
they implement, on the other hand, only simple concave 
prior densities corresponding to quadratic errors.
Especially complex tasks would require typically more general prior densities.
Choosing mixtures of Gaussian process priors
combines the advantage of an explicit formulation of priors
with the possibility of constructing
general non-concave prior densities.

While mixtures of Gaussian processes
are technically a relatively straightforward extension of Gaussian processes,
which turns out to be a computational advantage,
practically they are much more flexible
and are able to produce in principle, i.e.,
in the limit of infinite number of components,
any arbitrary prior density.

As example, consider an image completion task,
where an image have to be completed, given
a subset of pixels (`training data').
Simply requiring smoothness of grey level values
would obviously not be sufficient
if we expect, say, 
the image of a face.
In that case the prior density should reflect 
that a face has specific constituents (e.g., eyes, mouth, nose)
and relations (e.g., typical distances between eyes)
which may appear in various variations
(scaled, translated, deformed, varying lightening conditions).

While ways how prior mixtures can be used in such situations
have already been outlined in
\cite{Lemm-1996,Lemm-1998a,Lemm-1998b,Lemm-1999a,Lemm-Uhlig-Weiguny-1999}
this paper concentrates on the general formalism and
technical aspects of mixture models
and aims in showing their computational feasibility. 
Sections \ref{model}--\ref{prior-mixtures}
provide the necessary formulae
while Section \ref{numerical-ex} 
exemplifies the approach
for an image completion task.

Finally, we remark that mixtures of Gaussian process priors 
do usually {\it not} result in
a (finite) mixture of Gaussians
\cite{Everitt-Hand-1981} for the function of interest.
Indeed, in density estimation, for example, arbitrary densities
not restricted to a (finite) mixture of Gaussians
can be produced by a mixture of Gaussian prior processes.

\section{The Bayesian model}
\label{model}

Let us consider
the following random variables:
\begin{itemize}
\item[1.] 
$x$, representing (a vector of) 
{\it independent, visible variables} (`measurement situations'), 
\item[2.] 
$y$, being (a vector of) 
{\it dependent, visible variables} (`measurement results'),
and 
\item[3.] 
$h$, being the {\it hidden variables} (`possible states of Nature').
\end{itemize}
A Bayesian approach is based on two model inputs
\cite{Berger-1980,Robert-1994,Gelman-Carlin-Stern-Rubin-1995,Sivia-1996}:
\begin{itemize}
\item[1.]
A {\it likelihood model} $p(y|x,h)$,
describing the density of observing $y$ given $x$ and $h$.
Regarded as function of $h$, for fixed $y$ and $x$,
the density $p(y|x,h)$ 
is also known as the ($x$--conditional) {\it likelihood} of $h$.
\item[2.] 
A {\it prior model} $p(h|D_0)$, specifying
the {\it a priori} density of $h$ 
given some {\it a priori} information denoted by $D_0$
(but before training data $D_T$ have been taken into account).
\end{itemize}

\begin{figure}
\begin{center}
\setlength{\unitlength}{1mm}
\begin{picture}(60,50)
\put(0,0){\framebox(60,50)[]{}}
\put(20,46){\makebox(0,0){training data $D_T$}} 
\put(50,46){\makebox(0,0){test data}} 
\put(10,40){\makebox(0,0){$x_1$}}
\put(9.5,40){\circle{6}}
\put(20,40){\makebox(0,0){$\cdots$}} 
\put(30,40){\makebox(0,0){$x_n$}}
\put(29.5,40){\circle{6}}
\put(50,40){\makebox(0,0){$x$}}
\put(9.5,37){\vector(0,-1){4.5}}  
\put(29.5,37){\vector(0,-1){4.5}}  
\put(50,37){\vector(0,-1){4.5}}  
\put(10,30){\makebox(0,0){$y_1$}}
\put(9.5,30){\circle{6}}
\put(20,30){\makebox(0,0){$\cdots$}} 
\put(30,30){\makebox(0,0){$y_n$}}
\put(29.5,30){\circle{6}}
\put(50,30){\makebox(0,0){$y$}}
\put(37,23){\vector(-1,1){4.5}}  
\put(43,23){\vector(1,1){5}}   
\put(36,21){\vector(-4,1){20}}  
\put(41,20){\makebox(0,0){$(\beta),h$}}
\put(40,13){\vector(0,1){4}}  
\put(40,10){\makebox(0,0){$\theta, j$}}
\put(35,5){\dashbox{}(15,20)[br]{$\tilde {h}$ }}
\put(20,16){\makebox(0,0){prior data}} 
\put(23.2,10){\vector(1,0){10}}  
\put(20,10){\makebox(0,0){$D_0$}}
\put(20,10){\circle{6}}
\end{picture}
\end{center}
\caption{
Graphical representation of
the considered probabilistic model,
factorising according to
$p(x_T,y_T,x,y,h,\theta,j,(\beta)|D)$
=
$p(x_T)$ $p(x)$ 
$p(y_T|x_T,h,(\beta))$
$p(y|x,h,(\beta))$
$p(h|\theta,j,D_0,(\beta))$
$p(\theta,j,(\beta)|D_0)$.
(The variable $\beta$ is introduced
in Section \ref{regression}.)
Circles indicate visible variables.
}
\label{graph-model}
\end{figure}
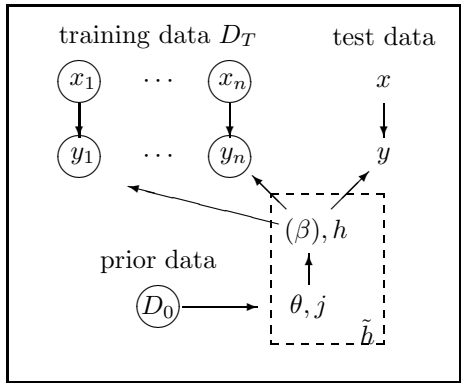

Furthermore, 
to decompose a possibly complicated {\it prior density}
into simpler components, 
we introduce
{\it continuous hyperparameters} $\theta$ 
and {\it discrete hyperparameters} $j$ 
(extending the set of hidden variables to $\tilde h$ = $(h,\theta,j)$),
\be
p(h|D_0) = \int \!d\theta \sum_j p(h,\theta,j|D_0)
.
\ee
In the following, the summation
over $j$ will be treated exactly,
while the $\theta$--integral will 
be approximated.
A Bayesian approach aims in calculating the {\it predictive density}
for outcomes $y$ in {\it test} situations  $x$
\be
p(y|x,D) = \int \!dh\, p(y|x,h)\, p(h|D)
,
\ee
given data $D$ = $\{D_T,D_0\}$
consisting of 
{\it a priori} data $D_0$
and
i.i.d.\ training data $D_T$ 
= $\{(x_i,y_i)|1\le i\le n\}$.
The vector of all $x_i$ ($y_i$)
will be denoted $x_T$ $(y_T)$.
Fig.\ref{graph-model} shows a graphical representation 
of the considered probabilistic model.

In saddle point approximation 
({\it maximum a posteriori approximation})
the $h$--integral becomes
\be
p(y|x,D) \approx p(y|x,h^*)
,
\ee
\be
h^*=\,{\rm argmax}_{h\in{\cal H}} p(h|D)
,
\label{maxi}
\ee
assuming $p(y|x,h)$ 
to be slowly varying at the stationary point.
The {\it posterior density} is related to 
($x_T$--conditional) likelihood and prior
according to Bayes' theorem
\be
p(h|D)
=\frac{p(y_T|x_T,h)\,p(h|D_0)}{p(y_T|x_T,D_0)}
\label{bayes-rule}
,
\ee
where the $h$--independent denominator (evidence)
can be skipped  when maximising with respect to $h$.
Treating the $\theta$--integral within $p(h|D)$
also in saddle point approximation
the posterior must be maximised
with respect to $h$ and $\theta$ simultaneously .

\section{Gaussian regression}
\label{regression}

In general density estimation problems 
$p(y_i|x_i,h)$ is not restricted to a special form,
provided it is non--negative and normalised
\cite{Lemm-1999a,Lemm-Uhlig-Weiguny-1999}.
In this paper we concentrate on Gaussian regression
where the single data likelihoods are assumed to be Gaussians
\be
p(y_i|x_i,h) = 
\sqrt{\frac{\beta}{2\pi}} e^{-\frac{\beta}{2} (h(x_i)-y_i)^2}
.
\label{gausslikeli}
\ee
In that case the 
unknown regression function $h(x)$
represents the hidden variables
and $h$--integration means
functional integration
$\int dh \rightarrow \int \prod_x dh(x)$.

As simple building blocks for mixture priors
we choose Gaussian (process) prior components
\cite{Doob-1953,Wahba-1990,Williams-Rasmussen-1996},
\bea&&
p(h|\beta, \theta, j, D_0) 
=
\left(\frac{\beta}{2\pi}\right)^{\frac{d}{2}}
\left(\det {\bf K}_j (\theta) \right)^\frac{1}{2}
\nonumber\\&&
\times
e^{-\frac{\beta}{2} \rscp{h-t_j(\theta)}{{\bf K}_j(\theta ) (h-t_j(\theta)) }}
\label{gaussprior}
\eea
the scalar product notation $\rscp{\cdot}{\cdot}$
standing for $x$--integration.
The mean $t_j(\theta)(x)$
will in the following 
also be called an (adaptive) {\it template function}.
Covariances ${\bf K}^{-1}_{j}/\beta$
are real, symmetric, positive (semi--)definite
(for positive semidefinite covariances the null space has to be projected out).
The dimension $d$ of the $h$--integral
becomes infinite for an infinite number of $x$--values
(e.g.\ continuous $x$).
The infinite factors appearing thus 
in numerator and denominator of (\ref{bayes-rule})
however cancel. 
Common smoothness priors
have $t_j (\theta)=0$
and as ${\bf K}_j$ a differential operator,
e.g., the negative Laplacian.

Analogously to simulated annealing
it will appear to be very useful to vary
the `inverse temperature' $\beta$
simultaneously
in (\ref{gausslikeli}) (for training but not necessarily for test data)
and (\ref{gaussprior}).
Treating $\beta$ not as a fixed variable,
but including it explicitly as hidden variable,
the formulae of Sect.\ \ref{model} remain valid, 
provided the replacement $h\rightarrow (h,\beta)$
is made, e.g.\ 
$p(y_i|x_i,h)\rightarrow p(y_i|x_i,h,\beta)$
(see also Fig.\ref{graph-model}).

Typically, inverse prior covariances can be related to 
{\it approximate symmetries}.
For example, assume we expect the regression function to be 
approximately invariant under a permutation of its arguments
$h(x) \approx h(\sigma(x))$ with 
$\sigma$ denoting a permutation.
Defining an operator  ${\bf S}$ acting on $h$ according to
${\bf S}h(x) = h(\sigma(x))$, 
we can define a prior process with 
inverse covariance 
\be
{\bf K} = ({\bf I}-{\bf S})^T ({\bf I}-{\bf S})
,
\ee
with identity ${\bf I}$
and the superscript ${}^T$ denoting the transpose
of an operator.
The corresponding prior energy
\be
E_0 
= \frac{1}{2} \left( h, \,{\bf K}\,h\right)
= \frac{1}{2} \Big( (h-{\bf S})h,\, (h-{\bf S})h\Big)
,
\ee
is a measure of the deviation of $h$ from 
an exact symmetry under ${\bf S}$.
Similarly, 
we can consider a Lie group
${\bf S}$ = $e^{\theta{\bf s}}$
with ${\bf s}$ being the generator 
of the infinitesimal symmetry transformation.
In that case
a covariance 
\be
{\bf K} 
= \frac{1}{\theta^2}
  ({\bf I}-{\bf S}_{\rm inf})^T({\bf I}-{\bf S}_{\rm inf})
= {\bf s}^T{\bf s}
,
\ee
with prior energy
\be
E_0 
= \frac{1}{2} \left( {\bf s}h, \,{\bf s}h\right)
,
\ee
can be used to implement approximate invariance
under the infinitesimal symmetry transformation
${\bf S}_{\rm inf}$ = ${\bf I} + \theta{\bf s}$.
For appropriate boundary conditions,
a negative Laplacian ${\bf K}$ 
can thus be interpreted as 
enforcing approximate invariance under
infinitesimal translations, i.e., for 
${\bf s}$ = $\partial/\partial x$.

\section{Prior mixtures}
\label{prior-mixtures}
\subsection{General formalism}
Decomposed into components
the posterior density becomes
\bea
p(h,\beta|D) &\!\propto\!&\! \int\!d\theta\,\sum_j^m
p(y_T|x_T,h,\beta)
\label{mix-model}
\\&&\!\!\!\!\!\!\times\;
p(h|\beta,\theta,j,D_0)
\,p(\beta,\theta,j|D_0)
\nonumber
.
\eea
Writing probabilities in terms of energies,
including parameter dependent normalisation factors 
and skipping parameter independent factors yields
\bea
 p(y_T|x_T,h,\beta ) 
 &\propto& 
 e^{-\beta E_T+\frac{n}{2}\ln \beta}
\nonumber\\
p(h|\beta,\theta,j,D_0)
&=&
e^{-\beta E_{0,j}+\frac{d}{2}\ln \beta}
\\
&&\times 
 e^{\frac{1}{2} \ln \det {\bf K}_j (\theta)}
\nonumber
\\
 p(\beta,\theta,j|D_0)
 &\propto&
 e^{-E_{\theta,\beta,j}}
.
\nonumber
\eea
This defines 
hyperprior energies $E_{\theta,\beta,j}$,
prior energies $E_{0,j}$
(`quadratic concepts')
\be
E_{0,j}
= \frac{1}{2}
\Brscp{h-t_j (\theta)} {{\bf K}_j (\theta) (h-t_j (\theta,j) )}
,
\ee
(the generalisation
to a sum of quadratic terms $E_{0,j} =\sum_k E_{0,k,j}$
is straightforward)
and 
training or likelihood energy (training error) 
\be
E_T 
= 
\frac{1}{2} \sum_i^n (h(x_i)-y_i)^2
\ee
\[
= 
\frac{1}{2} \left( \Brscp{h-t_T}{{\bf K_T} (h-t_T)}
                         +\sum_i^{n} V_{T}(x_i) \right)
.
\]
The second line is a
`bias--variance' decomposition 
where
\be
t_T (x_i) = \sum_k^{n_{x_i}} \frac{y_k(x)}{n_{x_i}}
,
\ee
is the mean of the $n_{x_i}$ training data available for $x_i$,
and 
\be
V_T(x_i) = 
\sum_k^{n_{x_i}} \frac{y^2_k(x)}{n_{x_i}} -t_T^2 (x_i)
,
\ee
is the variance of $y_i$ values at $x_i$.
($V_i$ vanishes if every $x_i$ appears only once.)
The diagonal matrix ${\bf K}_T$ is 
restricted to the space of $x$ for which
training data are available and has matrix elements $n_x$.


\subsection{Maximum a posteriori approximation}

In general density estimation
the predictive density can only be calculated 
approximately, e.g.
in maximum a posteriori approximation
or by Monte Carlo methods.
For Gaussian regression, however the predictive density
of mixture models can be calculated exactly 
for given $\theta$ (and $\beta$).
This provides us with the 
opportunity to compare the simultaneous 
maximum posterior approximation
with respect to $h$ and $\theta$
with an analytical $h$--integration
followed by a 
maximum posterior approximation
with respect to $\theta$.

Maximising the posterior 
(with respect to $h$, $\theta$, and
possibly $\beta$)
is equivalent to minimising
the mixture energy 
(regularised error functional
\cite{Tikhonov-Arsenin-1977,Wahba-1990,Vapnik-1982,Vapnik-1998})
\be
E = -\ln \sum_j^m e^{ -E_j + c_j}
,
\label{adapmixregr}
\ee
with component energies
\be
E_j = \beta E_{h,j} +E_{\theta,\beta,j}
,\quad
E_{h,j} = E_{T} +E_{0,j},
\label{adapmixE1}
\ee
and
\be
c_j (\theta,\beta)
= \frac{1}{2}\ln \det {\bf K}_j (\theta )
+\frac{d+n}{2}\ln \beta
.
\ee

In a direct saddle point approximation with respect
to $h$ and $\theta$
stationarity equations
are obtained by setting the (functional)
derivatives with respect
to $h$ and $\theta$ to zero,
\bea
0
\!\!\!&=&\!\!\!\! 
 \sum_j^m \!
a_j
\Big({\bf K}_T (h-t_T)+\!{\bf K}_j (h-t_j)\Big)
,
\;\;\;\;\;\;\;
\label{regr1}
\\
0 \!\!\!&=&\!\!\!\! 
 \sum_j^m 
a_j
\Bigg( 
      \frac{\partial E_{j}}{\partial \theta} 
    -{\rm Tr}\, \left(
                      {\bf K}_j^{-1}\frac{\partial {\bf K}_j}{\partial \theta} 
                 \right)
\Bigg)
,
\label{regr2}
\eea
where
the derivatives with respect
to $\theta$ are matrices
if $\theta$ is a vector,
\bea
a_j &=& p(j|h,\theta,D_0)
\label{a-eq}\\ 
&=& \frac{e^{-\beta E_{0,j}-E_{\theta,\beta,j}+\frac{1}{2}\ln\det{\bf K}_j}}
{\sum_k^me^{-\beta E_{0,k}-E_{\theta,\beta,k}+\frac{1}{2}\ln\det{\bf K}_k}}
,
\nonumber
\eea
and 
\bea
\frac{\partial E_{j}}{\partial \theta} &=&
\frac{\partial E_{\theta,\beta,j}}{\partial \theta}
+
\beta\left( \frac{\partial t_j}{\partial \theta},\; 
        {\bf K}_j  (t_j-h)\right)\
\nonumber\\&&
\!\!\!
+\frac{\beta}{2} 
\Big((h-t_j),\, 
\frac{\partial {\bf K}_j }
     {\partial \theta}(h-t_j)\Big)
.
\eea
Eq.(\ref{regr1}) can be rewritten
\be
h
=
{\bf K}_a^{-1}
\left( {\bf K}_T t_T + \sum_l^m a_j {\bf K}_j t_j \right)
,
\label{h-eq}
\ee
with
\be
{\bf K}_a
= \left( {\bf K}_T + \sum_j^m a_j {\bf K}_j \right)
.
\ee
Due to the presence of $h$--dependent factors $a_j$,
Eq.(\ref{h-eq}) is  still a nonlinear equation for $h(x)$.
For the sake of simplicity we assumed a fixed $\beta$; 
it is no problem however to solve (\ref{regr1}) and (\ref{regr2})
simultaneously with an analogous
stationarity equation for $\beta$.

\subsection{Analytical solution}

The optimal regression function under squared--error loss
--- for Gaussian regression identical to the log--loss
of density estimation ---
is the predictive mean.
For mixture model (\ref{mix-model})
one finds, say for fixed $\beta$,

\be
\bar y 
= \int\!dy\, y\, p(y|x,D)
= \sum_j  
\int\! d\theta \; b_j (\theta )\, \bar t_j (\theta)
,
\label{exact-mix-mean}
\ee
with 
mixture coefficients
\bea
b_j(\theta)
&=& p(\theta,j|D) 
\label{exact-mix-coeff}\\
&=&
\frac{p(\theta,j) \,
p(y_T|x_T,D_0,\theta,j)} 
{\sum_j\int \!d\theta p(\theta,j)\, p(y_T|x_T,D_0,\theta,j)} 
.
\nonumber
\eea
The component means $\bar t_j$
and
the likelihood of $\theta$ can be calculated analytically
\cite{Wahba-1990,Williams-Rasmussen-1996}
\bea
\bar t_{j}
&=&
\left( {\bf K}_T + {\bf K}_{j} \right)^{-1}
\left( {\bf K}_T t_T + {\bf K}_j t_{j} \right)
\nonumber\\
&=& t_j + {\bf K}_j^{-1} \widetilde {\bf K}_j (t_T-t_j)
\label{componentT}
,
\eea
and
\be
p(y_T|x_T,D_0,\theta,j)
=
e^{-\beta \widetilde E_{0,j}
+\frac{1}{2}\ln \det (\frac{\beta}{2\pi}\widetilde {\bf K}_j )}
\label{exact-theta-likeli}
,
\ee
where
\bea
\widetilde E_{0,j}(\theta) &=&
\frac{1}{2}  
\big( t_T - t_j ,\, 
    \widetilde {\bf K}_{j} (t_T - t_j )\big)
,\quad
\\
\widetilde {\bf K}_{j}(\theta)
&=& 
({\bf K}_T^{-1}+{\bf K}_{j,TT}^{-1})^{-1}
,
\label{Ktilde}
\eea
and ${\bf K}_{j,TT}^{-1}$ is the projection
of the covariance ${\bf K}_{j}^{-1}$ into the 
$\tilde n$--dimensional space
for which training data are available.
($\tilde n\le n$ is the number of data with distinct
$x$--values.)

The stationarity equation
for a maximum a posteriori approximation
with respect to $\theta$ 
is at this stage found from 
(\ref{exact-mix-coeff},\ref{exact-theta-likeli})
\be
0 = 
\sum_j b_j
\left(
\frac{\partial \widetilde E_{j}}{\partial \theta}
-{\rm Tr}\left(\widetilde {\bf K}_j^{-1}
  \frac{\partial \widetilde {\bf K}_{j}}{\partial \theta}\right)
\right),
\label{regr2a}
\ee
where
$\widetilde E_{j}$ = $\beta\widetilde E_{0,j}$ 
+ $E_{\theta,\beta,j}$.
Notice that Eq.(\ref{regr2a}) 
differs from Eq.(\ref{regr2}) 
and requires only to deal with 
the $\tilde n \times \tilde n$--matrix $\widetilde {\bf K}$.
The coefficient
$b^*_j$ = $b_j(\theta^*)$
for $\theta$ set to its maximum posterior value
is of form 
(\ref{a-eq})
with the replacements 
${\bf K}_j\rightarrow \widetilde{\bf K}_j $,
$E_j\rightarrow \widetilde E_j $.

\subsection{High and low temperature limits}

Low and high temperature limits
are extremely useful because in both cases
the stationarity Eq.(\ref{regr1}) becomes linear,
corresponding thus to 
classical quadratic regularisation approaches.

In the {\it high temperature limit}
$\beta \rightarrow 0$
the exponential factors $a_j$ become $h$--independent
\be
a_j 
\stackrel{\beta\rightarrow0}{\longrightarrow}
a^0_j =
\frac{e^{-E_{\theta,\beta,j}+\frac{1}{2}\ln\det{\bf K}_j}}
{\sum_k^me^{-E_{\theta,\beta,k}+\frac{1}{2}\ln\det{\bf K}_k}}
,
\label{highTa}
\ee
(for $b_j^{*}\rightarrow b_j^{0,*}$ 
replace ${\bf K}_j$ by $\widetilde{\bf K}_j $).
The solution 
$h = \bar t$
is a (generalised) `complete template average'
\be
\bar t =
{\bf K}_{a^0}^{-1}
\left( {\bf K_T} t_T 
+ \sum_l^m a^0_j \,{\bf K}_j t_j \right)
,
\label{completeT}
\ee
with 
\be
{\bf K}_{a^0} = {\bf K}_T+\sum_j a_j^0 \,{\bf K}_j 
.
\ee
This high temperature solution 
corresponds to the minimum of the quadratic
functional
$E_{T} +\sum_j^m a_j^0 E_{h,j}$,

In the {\it low temperature limit} $\beta\rightarrow \infty$
only the maximal component contributes,
i.e.,
\be
a_j \stackrel{\beta\rightarrow\infty}{\longrightarrow}
a^\infty_j =
\left\{ \begin{array}{r@{\quad : \quad}l}
1 & j = {\rm argmin}_j E_{h,j}
\\
0 & j \ne  {\rm argmin}_j E_{h,j}
\end{array} \right.
\label{lowTa}
,
\ee
(for $b_j^*$ 
replace $E_{h,j}$ by $\widetilde E_j $)
assuming $E_{\beta,\theta,j}$ = $E_{\beta}$ + $E_{\theta,j}$
or 
$E_{\beta,\theta,j}$ = $E_{\beta}$ + $E_{j}$ +$\beta E_{\theta}$.
Hence, 
low temperature solutions
$h = \bar t_{j}$,
are 
all (generalised) `component averages'  
$\bar t_j$
provided they
fulfil the stability condition 
\be
 E_{h,j} (h=\bar t_j)
<E_{h,j^\prime} (h=\bar t_j)
, \quad \forall j^\prime \ne j
,
\ee
or, after
performing a (generalised) `bias--variance' decomposition,
$2 V_{j}<{B}_{j^\prime} (j,j)  
 + 2 V_{j^\prime}$, 
with 
$m\times m$ matrices
\be
{B}_j (k,l) = 
\Big(\bar t_k-\bar t_j,\,\left( {\bf K}_D + {\bf K}_j\right)
\,(\bar t_l-\bar t_j)\Big)
\label{Bmatrix}
\ee
and (generalised) `template variances'
\bea
V_j \!\!&\!\!=& 
\frac{1}{2} \Bigg(
\Big(t_T,\,{\bf K}_T\,t_T\Big)
+\Big(t_j,\,{\bf K}_j\,t_j\Big)
\quad\nonumber\\&-&
\Big(\bar t_j,({\bf K}_T+{\bf K}_j) \,\bar t_j\Big)
\Bigg) = \widetilde E_{0,j}
\label{Vterm}
.
\eea
That means
single component averages
$\bar t_j$ 
(which minimise $E_{h,j}$ and thus $-\beta E_j+c_j$)
become solutions at zero temperature $1/\beta$
in case their (generalised) variance $V_j$ 
measuring the discrepancy between data and prior term is small
enough.

\subsection{Equal covariances}

Especially interesting 
are $j$--independent 
${\bf K}_j(\theta)$ = ${\bf K}_0 (\theta )$
with $\theta$--independent determinants so 
$\det {\bf K}_j$ or $\det \widetilde {\bf K}_j$, respectively, 
do not have to be calculated.

Notice that this still allows
completely arbitrary parameterisations 
of $t_j(\theta)$.
Thus, the template function can for example be 
a parameterised model,
e.g., a neural network or decision tree,
and maximising the posterior with respect to $\theta$ corresponds
to training that model.
In such cases the prior term
forces the maximum posterior solution $h$ to be similar
(as defined by ${\bf K_0}$)
to this trained parameterised reference model.

The condition of invariant
$\det{\bf K}_0(\theta)$ does not exclude 
adaption of covariances.
For example, transformations for real, symmetric positive definite 
${\bf K}_0(\theta)$ leaving 
determinant and eigenvalues (but not eigenvectors) invariant 
are of the form
${\bf K}(\theta_0)\rightarrow {\bf K}(\theta)=
{\bf O}(\theta){\bf K}{\bf O}^{-1}(\theta)$
with real, orthogonal 
${\bf O}^{-1}$ = ${\bf O}^{T}$.
This allows for example to adapt the sensible
directions of multidimensional Gaussians.
A second kind of transformations
changing eigenvalues but not eigenvectors and determinant
is of the form
${\bf K}(\theta_0) = {\bf O}{\bf D}(\theta_0){\bf O}^{T}$
$\rightarrow {\bf K}(\theta) = {\bf O}{\bf D}(\theta){\bf O}^{T}$
if the product of eigenvalues
of the real, diagonal 
${\bf D}(\theta_0)
$ and ${\bf D}(\theta)$ are equal.

\begin{figure}
\begin{center}
\setlength{\unitlength}{0.57mm}
\begin{picture}(44,36)
\put(0,0){\framebox(44,36)[]{}}
\put(22,26){\circle*{1.9}}
\put(23,30){\makebox(0,0){$\bar t_1$}}
\put(22,15){\circle*{1.9}}
\put(22,19){\makebox(0,0){$\bar t$}}
\put(15,12){\circle{1.9}}
\put(15,16){\makebox(0,0){$h$}}
\put(6,10){\circle*{1.9}}
\put(5,5){\makebox(0,0){$\bar t_2$}}
\put(38,10){\circle*{1.9}}
\put(37,5){\makebox(0,0){$\bar t_3$}}
\put(22.5,25.5){\line(1,-1){15}}  
\put(21.5,25.5){\line(-1,-1){15}}  
\put(6.8,10){\line(1,0){30.7}}  
\put(90,42){\makebox(0,0){$\beta$}}
\put(69,21){\makebox(0,0){$a_1$}}
\put(78,12){\makebox(0,0){$b_1$}}
\end{picture}
\hspace{0.2cm}
\includegraphics[scale = .37 ]{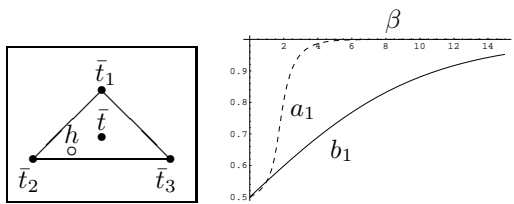}
\end{center}
\vspace{-0.5cm}
\caption{Left: Example of a solution space for $m$ = 3.
Shown are three low temperature solutions $\bar t_j$,
high temperature solution $\bar t$, and a possible
solution $h$ at finite $\beta$.
Right: Exact $b_1$ vs. (dominant) $a_1$ (dashed) 
for $m$ = $2$, $b$ = 2, 
$\widetilde E_1$ = 0.405,
$\widetilde E_2$ = 0.605.}
\label{m3-scheme}
\end{figure}

Eqs.(\ref{componentT},\ref{completeT})
show that the high temperature solution becomes
a linear combination
of the (potential) low temperature solutions
\be
\bar t 
= \sum_j^m a^0_j \bar t_j
= \sum_j^m b^{0,*}_j \bar t_j
.
\ee
Similarly,
Eq.(\ref{regr1}) simplifies to
\be
h 
=\sum_j^m a_j \bar t_j
=\bar t + \sum_j^m (a_j-a_j^0) \, \bar t_j
,
\label{equalcov}
\ee
and 
Eq.(\ref{a-eq}) to
\be
a_j= \frac{e^{-\frac{\beta}{2}a {B}_j a-\widetilde E_j}}
       {\sum_k e^{-\frac{\beta}{2} a {B}_k a-\widetilde E_k}}
= \frac{b_j \,e^{-\frac{\beta}{2}a {B}_j a}}
       {\sum_k b_k\, e^{-\frac{\beta}{2} a {B}_k a}}
,
\label{stata}
\ee
introducing
vector $a$ with components $a_j$,
$m\times m$ matrices $B_j$
defined in (\ref{Bmatrix}).
Eq.(\ref{equalcov}) is still a nonlinear equation
for $h$, it shows however that the solutions
must be convex combinations of the $h$--independent $\bar t_j$
(see Fig. \ref{m3-scheme}).
Thus, it is sufficient to solve
Eq.(\ref{stata}) for $m$ mixture coefficients $a_j$
instead of Eq.(\ref{regr1}) for the function $h$. 

\begin{figure}
\vspace{-1.5cm}
\begin{center}
$\!\!\!\!\!\!\!\!\!\!$
\includegraphics[scale = .35 ]{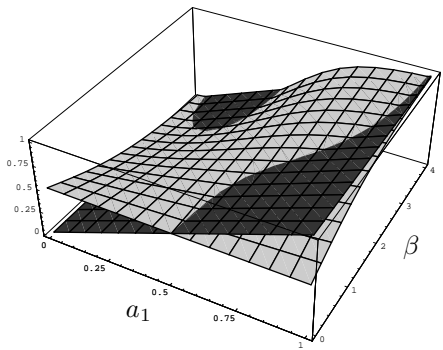}
\setlength{\unitlength}{1mm}
\end{center}
\begin{picture}(0,0)
\put(170,85){\makebox(0,0){$a_1$}}
\put(273,110){\makebox(0,0){$\beta$}}
\end{picture}
\vspace{-2.7cm}
\caption{
Shown are 
the plots of $f_1(a_1)=a_1$ 
and $f_2(a_1)=\frac{1}{2} \left(\tanh \Delta + 1\right)$
within the inverse temperature range $0\le\beta\le 4$
(for $b=2$, $\widetilde E_2-\widetilde E_1$ = $0.1\beta$).
Notice the appearance of a second stable
solution at low temperatures.}
\label{tpic}
\end{figure}

For two prior components, i.e., $m=2$, 
Eq.(\ref{equalcov}) becomes
\be
h 
=\frac{\bar t_1 + \bar t_2}{2} 
+ \left(\tanh \Delta\right)  \frac{\bar t_1 - \bar t_2}{2} 
,
\ee
with
\be
\Delta
 =
\frac{E_2-E_1}{2} 
=
\frac{\beta}{4} \,b(2 a_1-1) + \frac{\widetilde E_2-\widetilde E_1}{2}
,
\ee
because
the matrices $B_j$ are in this case zero except
$B_1(2,2) = B_2(1,1) = b$.
For $E_{\theta,\beta,j}$ uniform in $j$
we have
$(\bar t_1+\bar t_2)/2$ = $\bar t$ 
so that $a_j^0$ = $0.5$.
The stationarity Eq.(\ref{stata}),
being analogous to
the celebrated mean field equation of a ferromagnet,
can be solved graphically 
(see Fig.\ref{tpic} and Fig.\ref{m3-scheme}
for a comparison with $b_j$),
the solution is given by the point
where 
\be
a_1 = \frac{1}{2} \left(\tanh \Delta + 1\right)
.
\ee

\section{A numerical example}
\label{numerical-ex}

\begin{figure}
\begin{center}
\includegraphics[scale = 0.12]{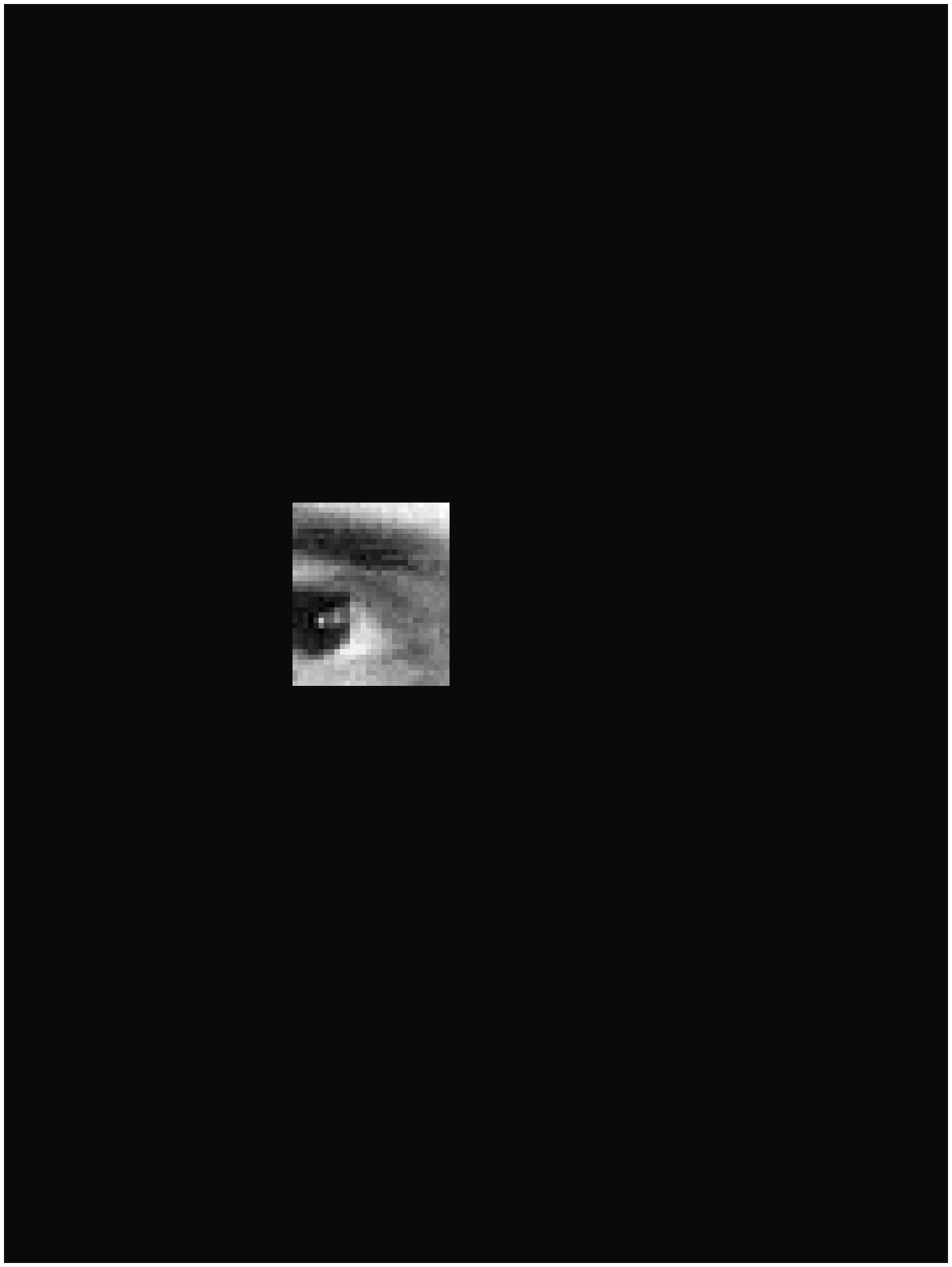}
\includegraphics[scale = 0.12]{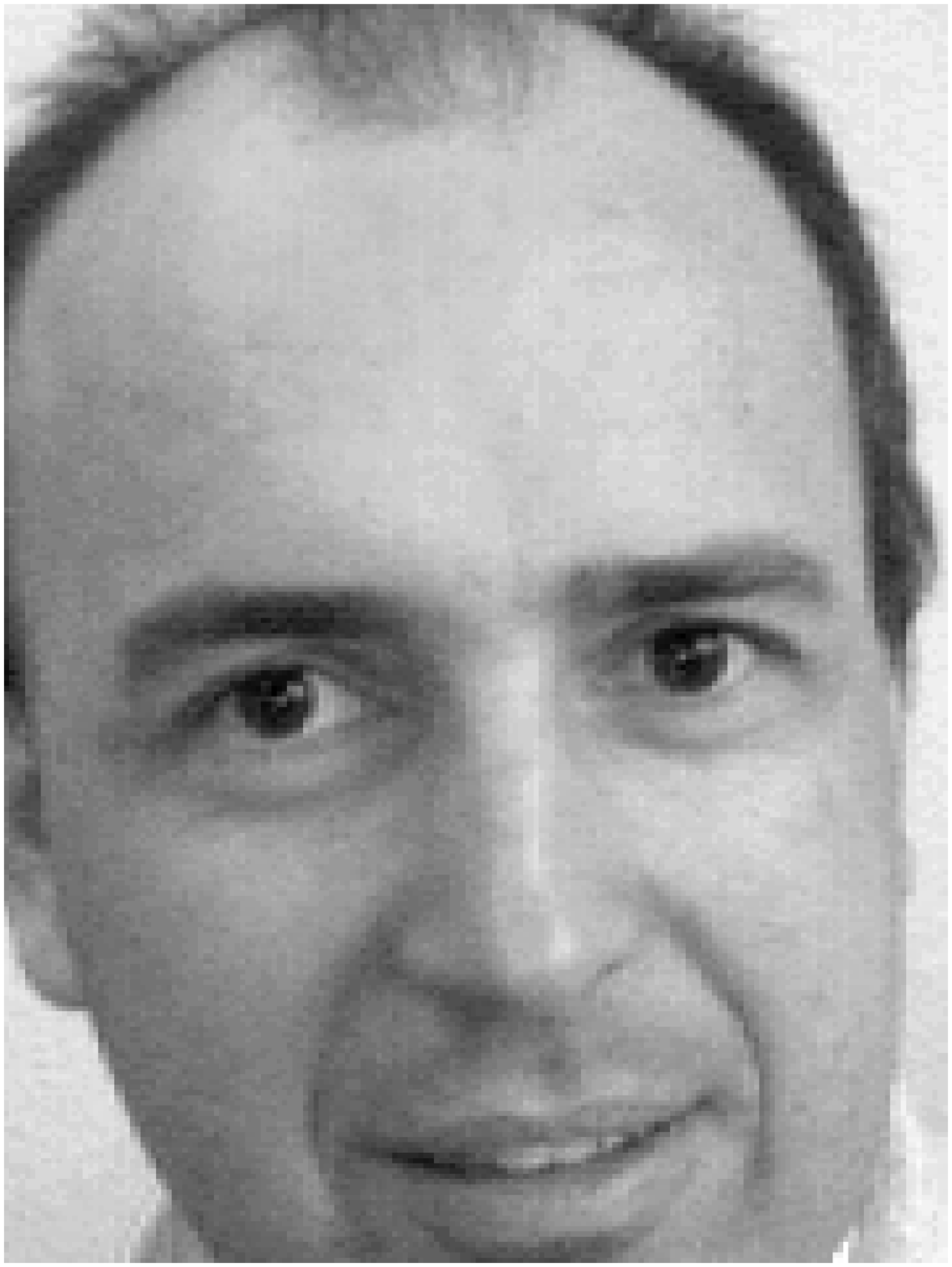}
\includegraphics[scale = 0.12]{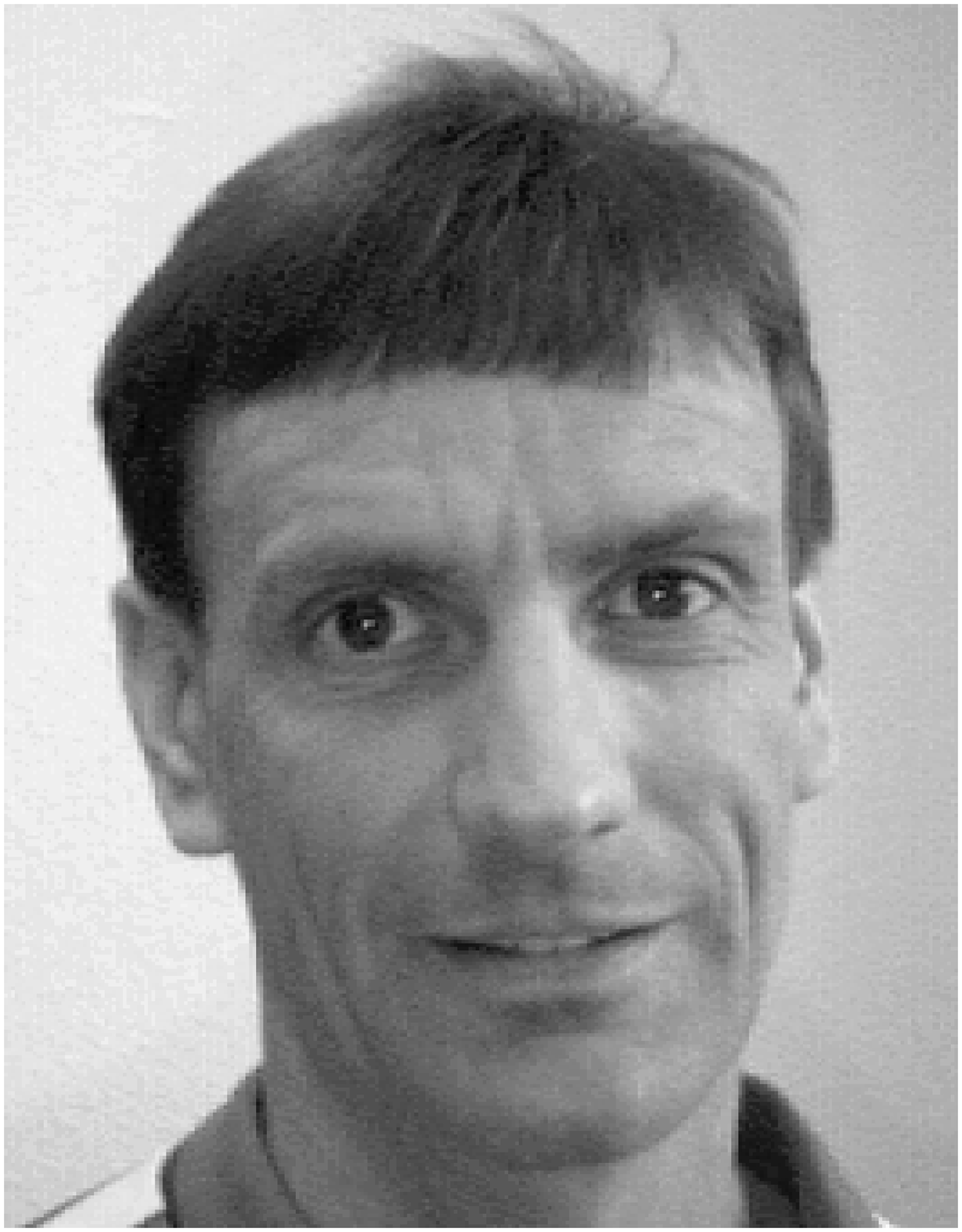}
\end{center}
\begin{center}
\includegraphics[scale = 0.12]{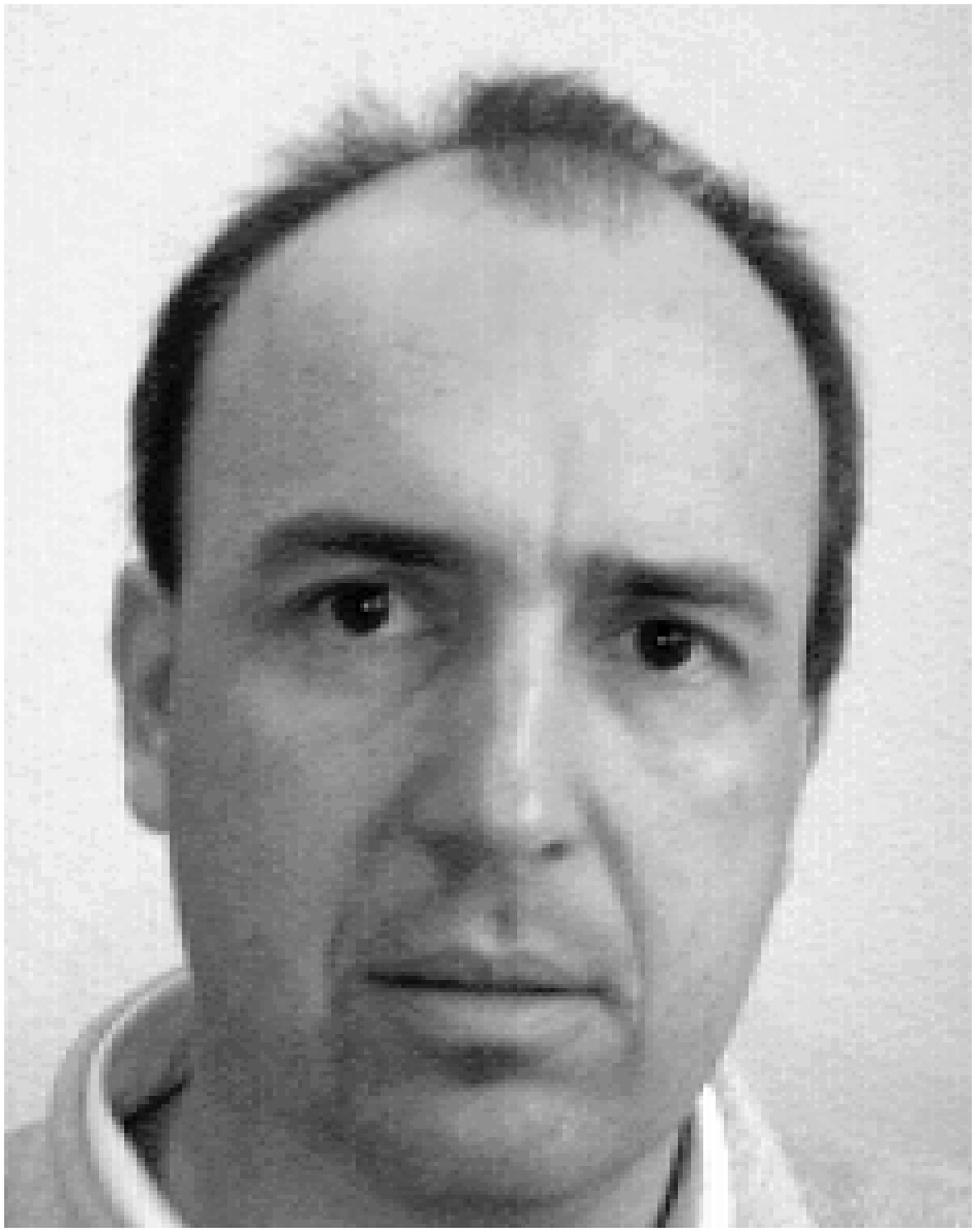}
\includegraphics[scale = 0.12]{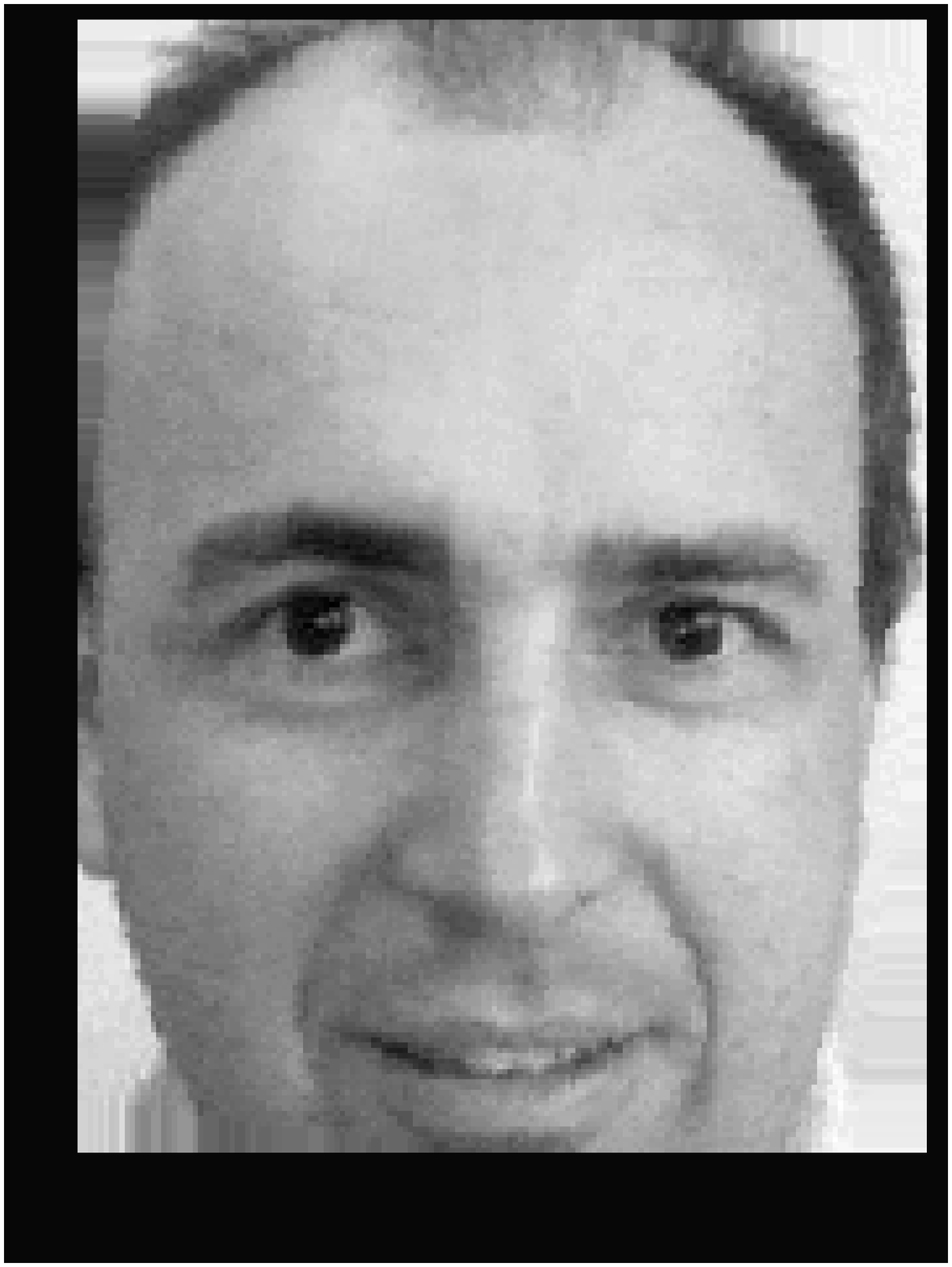}
\includegraphics[scale = 0.12]{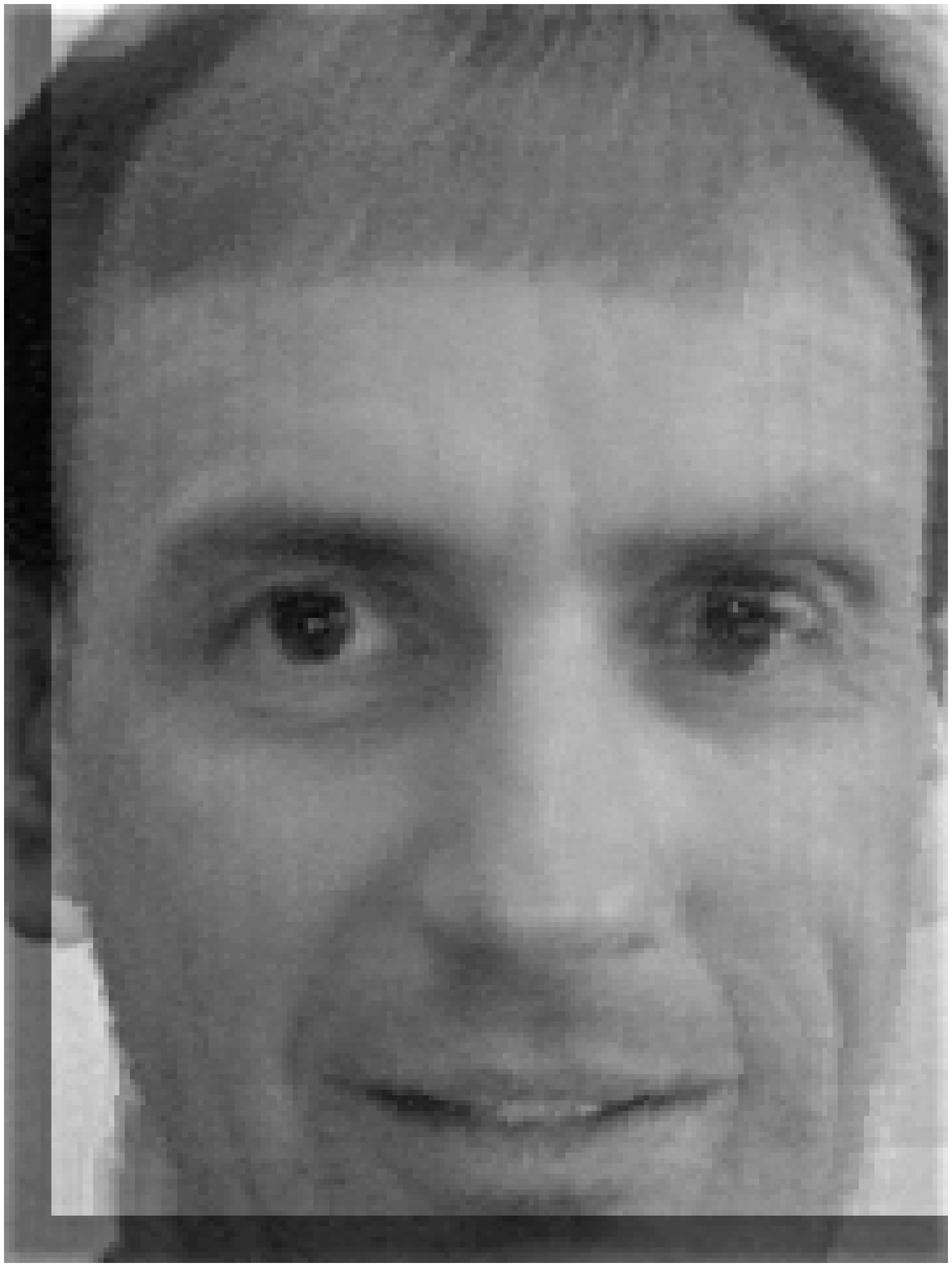}
\end{center}
\vspace{-0.5cm}
\caption{Top row, from left to right:
Data points sampled with Gaussian noise,
two template functions $t_1$, $t_2$.
Bottom row, from left to right:
Original, reconstructed solutions
(regression function $h$, 180$\times$240 pixels) at low and at
high temperature.
}
\label{images}
\vspace{-0.3cm}
\end{figure}

As numerical example we study a 
two component mixture model for image completion.
Assume we expect an only partially known image 
(corresponding to pixel-wise training data 
drawn  with Gaussian noise from the original image)
to be similar to one of the two template images shown 
in Fig.\ref{images}.
Next, we include hyperparameters parameterising deformations of templates.
In particular, we have chosen
translations ($\theta_1$, $\theta_2$)
a scaling factor $\theta_3$, 
and a rotation angle (around template center)
$\theta_4$.

Interestingly, 
it turned out that due to the large number of data 
($\tilde n\approx$ 1000) it was easier
to solve Eq.(\ref{regr1}) for the full discretized image
than to invert (\ref{Ktilde}) in the space of training data. 
A prior operator ${\bf K}_0$ has been implemented
as a $3\times3$ negative Laplacian filter.
(Notice that using a Laplacian kernel, or another smoothness measure,
instead of a straight template matching
using simply the squared error between image and template,
leads to a smooth interpolation between data and templates.)
Completed images $h$ 
for different $\beta$ have been found by iterating according to 
\be
h^{k+1} = h^k + \eta {\bf A}^{-1} 
\Big[
{\bf K}_T (t_T-h^k) 
+ {\bf K}_0 \Big(\sum_j a_j^k t_j -h^k\Big)
\Big]
,
\ee
performed alternating with $\theta$--minimisation.
A Gaussian learning matrix ${\bf A}^{-1}$ (implemented by
a $5\times 5$ binomial filter) proved to be successful. 
Typically, the relaxation factor $\eta$ has been set to $0.05$. 

Being a mixture model with $m=2$ 
the situation is that of Fig.\ref{tpic}.
Typical solutions for large and small $\beta$
are shown in Fig.\ref{images}.

\section{Conclusions}

Prior mixture models are capable to build complex prior densities
from simple, e.g., Gaussian components.
Going beyond classical quadratic regularisation approaches,
they still can use the nice analytical features of Gaussians,
and allow to control the degree of the resulting
non-convexity explicitly.
Combined with parameterised
component mean functions and covariances
they seem to provide a powerful tool.

\vspace{0.5cm}
\noindent{\bf Acknowledgements}
{\small 
The author was supported by a 
Postdoctoral Fellowship (Le 1014/1--1)
from the Deutsche Forschungsgemeinschaft and a NSF/CISE Postdoctoral
Fellowship at the Massachusetts Institute of Technology. 
Part of the work was done
during the
seminar `Statistical Physics of Neural Networks'
at the Max--Planck--Institut f\"ur Physik komplexer
Systeme, Dresden.
The author also wants to thank Federico Girosi, Tomaso Poggio, 
J\"org Uhlig, and Achim Weiguny for discussions.
}

\end{document}